\PassOptionsToPackage{hidelinks,colorlinks=true}{hyperref}
\documentclass[conference]{IEEEtran}
\IEEEoverridecommandlockouts

\usepackage{cite}
\usepackage{amsmath,amssymb,amsfonts}
\usepackage{algorithmic}
\usepackage{graphicx}
\usepackage{textcomp}
\usepackage{xcolor}
\usepackage{orcidlink}
\usepackage[version=4]{mhchem}
\usepackage{hyperref}

\hypersetup{
    linkcolor=blue,
    citecolor=blue,
    urlcolor=blue
}

\def\BibTeX{{\rm B\kern-.05em{\sc i\kern-.025em b}\kern-.08em
    T\kern-.1667em\lower.7ex\hbox{E}\kern-.125emX}}
\begin{document}

\title{Additive binding energies in asphalt on a quantum processor via quantum-selected configuration interaction (QSCI)}

\author{\IEEEauthorblockN{1\textsuperscript{st} Karim Elgammal\orcidlink{0000-0002-8222-3157}}
\IEEEauthorblockA{\textit{Researcher at RISE Research Institute of Sweden} \\
Stockholm, Sweden \\
karim.elgammal@ri.se}
\and
\IEEEauthorblockN{2\textsuperscript{nd} Marc Mau{\ss}ner}
\IEEEauthorblockA{\textit{Senior Consultant at infoteam Software AG}\\
Bubenreuth, Germany \\
marc.maussner@infoteam.de}
}

\maketitle

\begin{abstract}
Quantum-centric supercomputing (in which a quantum processor samples the dominant electronic configurations and classical high-performance computing resources perform the diagonalisation) is emerging as a practical route to correlated electronic-structure calculations. We present QuantumPave, a hybrid quantum-classical workflow for computing additive binding energies in asphalt binder, a quantity central to the oxidative ageing of road infrastructure. Using a 24-atom pyridine-phenol hydrogen-bonded complex as a representative model, we couple machine-learning interatomic potentials (ORB v3) for geometry optimisation with quantum-selected configuration interaction (QSCI), also referred to as sample-based quantum diagonalisation (SQD), in a (10e, 10o) active space run on the 54-qubit IQM Emerald processor. On hardware, SQD reproduces the active-space CASCI reference exactly, giving a binding energy of -3.52~kcal/mol (-0.153~eV); the device noise broadens the sampling to span the active space, so no zero-noise extrapolation is required. This active-space value captures the static correlation within the chosen orbitals and underbinds the full hydrogen bond: a counterpoise-corrected CCSD(T) benchmark gives -8.5 to -9.5~kcal/mol, while the calorimetric enthalpy of about -6.25~kcal/mol is consistent with this once zero-point, thermal, and solvent contributions are included. We show that chemically meaningful binding energies for an industrially relevant materials problem are attainable on current quantum hardware within a quantum-centric supercomputing workflow.
\end{abstract}

\begin{IEEEkeywords}
quantum-centric supercomputing, quantum-selected configuration interaction (QSCI), sample-based quantum diagonalisation (SQD), quantum computing, binding energy, density functional theory, machine-learning interatomic potentials, error mitigation, asphalt binder, HPC, Qiskit, IQM
\end{IEEEkeywords}
\section{Introduction}
\label{sec:introduction}

Asphalt binder is a complex organic material crucial for road infrastructure. Its bulk chemical composition governs long-term durability and ageing behaviour~\cite{petersen2000chemistry}, and molecular-dynamics simulations are increasingly used to relate this composition to oxidative ageing and binder--aggregate adhesion~\cite{XU2023,orear2025asphalt}. Their effectiveness depends on molecular interactions within the asphalt matrix which can be quantified using binding energy calculations, calculating the binding energy between an additive and key asphalt components is fundamental to understanding how additives interfere with ageing pathways \cite{qu2018aging}. This quantity directly relates to the additive's ability to protect asphalt from oxidation. We aim to build a model of asphalt binder ageing using a hybrid quantum classical approach. We focus on developing and applying a quantum-centric workflow to compute these additive binding energies. We aim to investigate the quantum algorithms benchmarking accuracy for the interaction energies within a computationally feasible active space, offering a proof-of-concept employing quantum and classical resources.

The accurate prediction of binding energies between molecular components remains a significant challenge in computational chemistry, particularly when strong correlation effects and dispersion interactions play crucial roles. Recent advances in quantum computing have opened new avenues for addressing these challenges through quantum-inspired algorithms that can capture many-body correlation effects beyond mean-field approximations \cite{McArdle2020,bauer2020review,robledomoreno2024}. The simulation of molecular energies on quantum hardware was proposed two decades ago~\cite{AspuruGuzik2005}; since then, hardware-efficient variational eigensolvers have been demonstrated for small molecules~\cite{kandala2017hardware}, and quantum algorithms have begun to address genuine chemical problems such as reaction mechanisms~\cite{diels_alder_quantum2024}. Although today's demonstrations operate in the noisy intermediate-scale quantum (NISQ) regime, the field is transitioning towards early fault-tolerant, so-called ``megaquop'', devices~\cite{preskill2025megaquop}, with error-mitigation techniques bridging the two regimes. The phenol-pyridine system represents an ideal test case for this methodology, as it exhibits both strong hydrogen bonding and $\pi$-conjugated electron systems. These interactions are particularly relevant in adhesion between asphalt and aggregate binder studies \cite{XU2023}.

Experimental studies have established that pyridine interactions with phenolic compounds in aqueous solutions form stable 1:1 hydrogen-bonded complexes with equilibrium constants of 0.6-0.7 M$^{-1}$ \cite{pekary1978}. The pyridine heterocycle is endowed with unique physicochemical properties, including weak basicity, high dipole moment that supports $\pi$-$\pi$ stacking interactions, and robust dual hydrogen-bonding capacity that could be important for asphalt components.

These methods sit within the emerging paradigm of \emph{quantum-centric supercomputing}, in which a quantum processor samples the dominant electronic configurations while classical high-performance computing resources perform the diagonalisation~\cite{robledomoreno2024}. Over the past year this paradigm has been formalised in a reference architecture for integrating QPUs with CPUs and GPUs~\cite{seelam2026qcsc}, and has scaled quantum--classical electronic-structure calculations of protein--ligand complexes beyond twelve thousand atoms on leadership-class supercomputers~\cite{merz2026protein}; the present work instantiates it at a deliberately small, fully reproducible scale, extending our earlier quantum-centric study of corrosion inhibition~\cite{elgammal2024} to asphalt binder chemistry. In this work, we present the full execution and validation of our approach naming it QuantumPave, demonstrating the practical application of quantum-selected configuration interaction (QSCI)~\cite{kanno2023qsci}, also referred to as sample-based quantum diagonalisation (SQD), to binding energy calculations \cite{Barison_2025,yu2025quantum}. QSCI continues to find new applications, recently including the photochemistry of small organic compounds~\cite{shivpuje2025sqd} and molecular complexes such as the pyridine-lithium ion system~\cite{ghasemi2026pyridineli}. We benchmark the quantum-selected configuration interaction result against its classical active-space reference and the experimental hydrogen-bond enthalpy, with machine-learning interatomic potentials and density functional theory supporting the geometry optimisation and providing classical context.
\section{Methodology}
\label{sec:methodology}

\subsection{System Selection}

We selected the pyridine-phenol complex (C$_{11}$H$_{11}$NO) containing 24 atoms as our system. This molecular complex was chosen for several reasons: firstly, it exhibits strong non-covalent interactions including hydrogen bonding and $\pi$-$\pi$ stacking that takes place experimentally; secondly, the system size remains computationally visible for quantum-inspired approaches and computational power available for simulation; thirdly, experimental and theoretical literature exist; and finally, it represents a realistic system for asphalt binder interactions. In particular, the phenolic hydroxyl and aromatic-nitrogen moieties of the complex mirror the oxygen- and nitrogen-bearing functional groups that recent molecular-dynamics studies identify as central to binder ageing and adhesion~\cite{orear2025asphalt}.

\begin{figure*}[h!]
\centering
\includegraphics[width=0.6\textwidth]{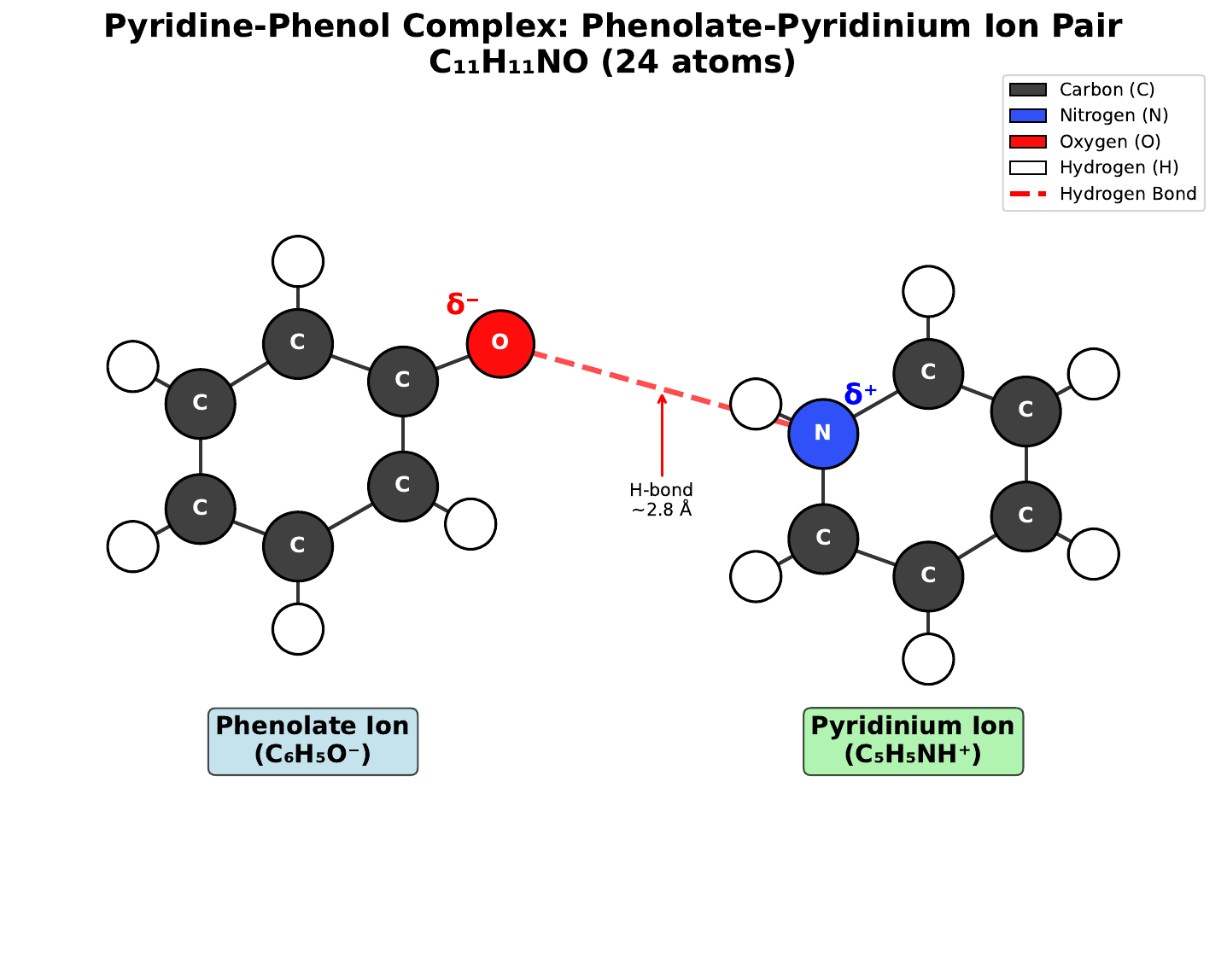}
\caption{Molecular structure of the pyridine-phenol complex. The system consists of 24 atoms (C$_{11}$H$_{11}$NO) and is held together by an O-H$\cdots$N hydrogen bond between the phenol hydroxyl group and the pyridine nitrogen.}
\label{fig:molecular_structure}
\end{figure*}

\begin{figure*}[h!]
\centering
\includegraphics[width=0.5\textwidth]{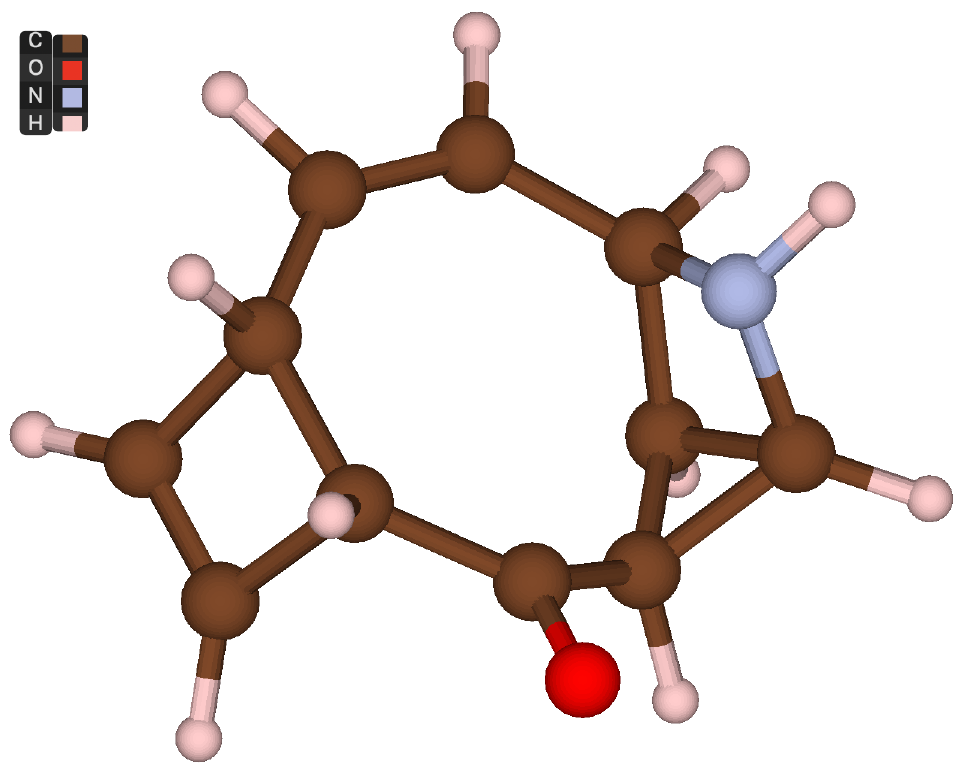}
\caption{Three-dimensional visualisation of the optimised pyridine-phenol complex structure obtained from ORB v3 calculations (Energy: -148.936 eV, 103 optimisation steps). The structure shows the hydrogen bonding interaction between the nitrogen in pyridine and oxygen in phenol, with molecular dimensions of 5.5 × 6.3 × 4.1 Å.}
\label{fig:supercell_3d}
\end{figure*}

The complex is held together by an O-H$\cdots$N hydrogen bond between the phenol hydroxyl group and the pyridine nitrogen. The SMILES representation of the two neutral fragments is C1=CC=C(C=C1)O.C1=CC=NC=C1, and the active space is constructed from the relaxed neutral pyridine, phenol, and complex geometries. Fig~\ref{fig:molecular_structure} illustrates the different interactions taking place. The system we study is depicted in Fig.~\ref{fig:supercell_3d} showing the pyridine-phenol complex.

\subsection{Computational Approaches}

Three complementary computational methodologies were implemented to provide an assessment of binding energies and validate our quantum-inspired approach: machine-learning interatomic potentials (ORB v3), dispersion-corrected density functional theory, and QSCI.

\subsubsection{ORB v3 Machine Learning Models}

The ORB v3 methodology employs machine learning potentials trained on extensive quantum chemistry datasets \cite{orb_fast_scalable2024, rhodes2025orbv3}. These models provide rapid energy predictions by leveraging pre-computed electronic structure information, offering a balance between computational efficiency and chemical accuracy. The ORB v3 conservative model with infinite neighbours was utilised for all single-point energy calculations, providing binding energy estimates within seconds of computation time, with a promised quality nearing highly accurate methods and with dispersion correction. Recent systematic benchmarks confirm that such foundational potentials, particularly once fine-tuned, approach \emph{ab initio} accuracy across diverse architectures~\cite{hanseroth2025finetuning}, supporting their use as an efficient reference in this work.

\subsubsection{Classical Reference Calculations}

Classical reference values were computed in PySCF. A supermolecular B3LYP/6-31+G(d) interaction energy provides a density-functional cross-check. The correlated gold standard is a counterpoise-corrected CCSD(T) calculation carried out on the LUMI supercomputer: directly in the def2-SVP basis, and at the def2-TZVP level through a focal-point estimate that combines MP2/def2-TZVP with the def2-SVP CCSD(T) triples correction, which for a hydrogen bond is accurate to about 0.1 to 0.2~kcal/mol. Counterpoise correction~\cite{boys1970calculation} removes basis-set superposition error in the supermolecular interaction energies, and the dispersion interactions central to such non-covalent binding are recovered by the correlated treatment itself~\cite{grimme2010consistent}.

\subsubsection{Quantum-Selected Configuration Interaction}

The core of this work is our quantum-inspired methodology, quantum-selected configuration interaction (QSCI)~\cite{kanno2023qsci}, also referred to as Sample-based Quantum Diagonalisation (SQD)~\cite{robledomoreno2024}. The approach employs active space approximations to reduce the computational complexity whilst target maintaining chemical accuracy for valence-dominated interactions. Active space selection utilised an automated frozen core approach that identifies core orbitals and focuses the active space on valence orbitals participating in intermolecular bonding \cite{qicas2023}.

\begin{figure*}[!ht]
\centering
\includegraphics[width=1.0\textwidth]{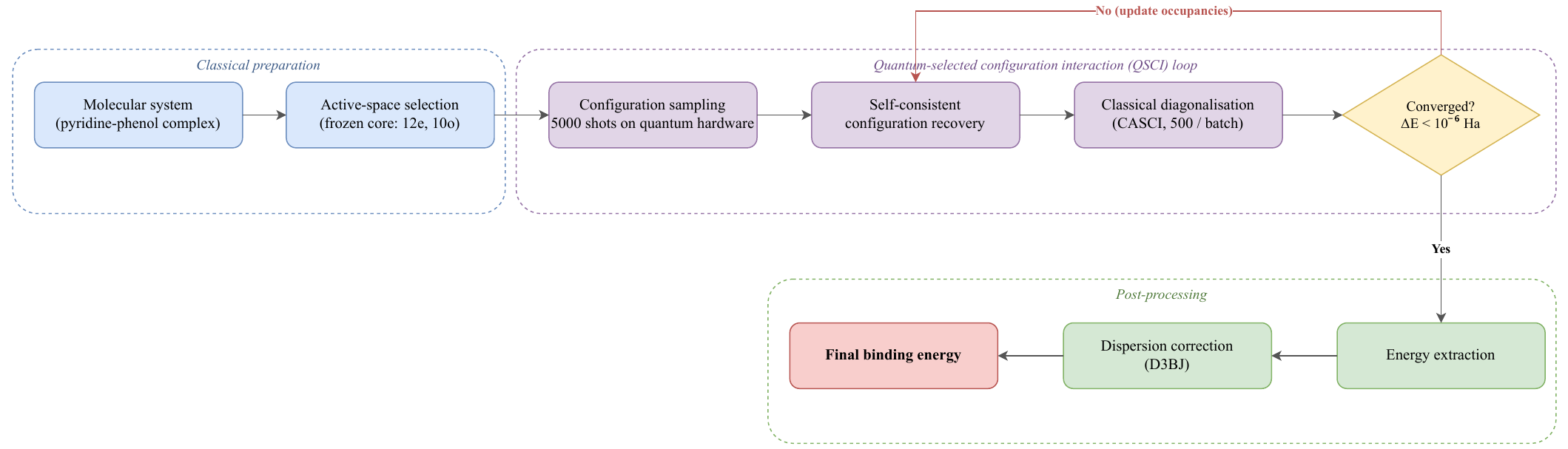}
\caption{QSCI workflow showing the active space selection, configuration sampling, and energy convergence procedures.}
\label{fig:sqd_architecture}
\end{figure*}

The QSCI algorithm is used as implemented in the standard format~\cite{robledomoreno2024, qiskit-addon-sqd, qiskit_sqd_docs}. On noisy hardware, the bitstrings sampled from the prepared state are subject to errors that violate particle-number symmetry; to counter this, the algorithm applies \emph{self-consistent configuration recovery}, which iteratively repairs invalid configurations using the average orbital occupations of the retained samples before each diagonalisation~\cite{robledomoreno2024}. The QSCI implementation incorporated several key parameters: active space configuration comprising 10 active electrons in 10 active orbitals (20 qubits) using Complete Active Space Configuration Interaction (CASCI); sampling parameters including 5000 measurement shots with 500 samples per batch; convergence criteria set to $10^{-6}$ Ha energy tolerance; and the 6-31G basis set for electronic structure calculations. The full active space spans $\binom{10}{5}^2 = 63\,504$ configurations, which remains tractable for the classical diagonalisation. The input state is a LUCJ/UCJ ansatz built from CCSD $t_1$ and $t_2$ amplitudes. The hardware run uses no zero-noise extrapolation: the self-consistent configuration recovery, together with the device noise that broadens the sampled distribution, allows the retained subspace to span the active space. The reported active-space binding energy is the bare CASCI value within the chosen orbitals, without an added empirical dispersion correction; the dynamic-correlation and dispersion contributions that lie outside the active space are quantified separately by the CCSD(T) reference. Fig~\ref{fig:sqd_architecture} shows those steps.

\subsection{Binding Energy Calculations}

Binding energies were calculated using the supermolecular approach for all three methodologies:

\begin{equation}
E_{\text{binding}} = E_{\text{complex}} - E_{\text{pyridine}} - E_{\text{phenol}}
\end{equation}

where $E_{\text{complex}}$, $E_{\text{pyridine}}$, and $E_{\text{phenol}}$ represent the total energies of the complex and isolated constituent molecules, respectively. The supermolecular approach places demands on the size consistency of the underlying solver; a size-consistent implementation of QSCI for intermolecular interaction energies has recently been developed~\cite{sugisaki2025sizeconsistent}.
\section{Results and Discussion}
\label{sec:results}

\subsection{Binding Energy Comparison}

Table~\ref{tab:binding_energies} presents the binding energy of the pyridine-phenol complex obtained on quantum hardware, alongside the classical active-space reference and the available experimental value.

\begin{table}[h!]
\centering
\caption{Binding Energy of the Pyridine-Phenol Complex}
\label{tab:binding_energies}
\resizebox{\columnwidth}{!}{%
\begin{tabular}{|l|l|c|c|}
\hline
\textbf{Method} & \textbf{Backend} & \textbf{kcal/mol} & \textbf{eV} \\
\hline
SQD & IQM Emerald (hardware) & -3.52 & -0.153 \\
\hline
CASCI & active space (reference) & -3.52 & -0.153 \\
\hline
CCSD(T)/def2-SVP+CP & classical (LUMI) & -8.53 & -0.370 \\
\hline
CCSD(T)/def2-TZVP+CP & focal-point (LUMI) & -9.51 & -0.412 \\
\hline
B3LYP/6-31+G(d) & supermolecular & -9.08 & -0.394 \\
\hline
experiment ($\Delta H$) & calorimetry (CCl$_4$) & -6.2 to -6.3 & $\approx -0.27$ \\
\hline
\end{tabular}%
}
\end{table}

SQD on the IQM Emerald processor reproduces the active-space CASCI reference exactly, both giving a binding energy of -3.52~kcal/mol (-0.153~eV). This active-space value captures the static correlation within the ten chosen valence orbitals but, by construction, omits the dynamic correlation and electrostatic contributions from the wider orbital space on which a weak, electrostatically dominated O-H$\cdots$N hydrogen bond predominantly depends. It therefore underbinds the full interaction energy: the counterpoise-corrected~\cite{boys1970calculation} CCSD(T) gold standard gives -8.53~kcal/mol in the def2-SVP basis and -9.51~kcal/mol at the focal-point def2-TZVP level (Table~\ref{tab:binding_energies}).

\subsection{Computational Accuracy and Method Agreement}

The central outcome is that SQD on real hardware reproduces the CASCI reference to numerical precision, the two agreeing to within $6\times10^{-8}$~Ha for the complex and both fragments. Because QSCI restricts the quantum processor to sampling and performs the diagonalisation classically within the selected subspace, the estimated energy obeys the variational principle and is insensitive to the statistical and hardware errors that would otherwise bias an expectation-value estimate. On IQM Emerald the noisy samples spanned essentially the full active-space configuration space, so the classical diagonalisation recovered the active-space energy exactly for the complex and both isolated molecules.

The CCSD(T) series is well behaved for this hydrogen bond, the binding energy evolving from Hartree-Fock (-6.78~kcal/mol, without dispersion) through MP2 (-9.19) and CCSD (-8.07) to CCSD(T) (-8.53) in the def2-SVP basis, and growing in magnitude towards the complete-basis limit. A supermolecular B3LYP/6-31+G(d) calculation (-9.08~kcal/mol) is consistent with this ladder. These are gas-phase electronic interaction energies; the calorimetric enthalpy of -6.2 to -6.3~kcal/mol measured for the 1:1 complex in carbon tetrachloride~\cite{spencer1987phenolpyridine} is smaller largely because of zero-point, thermal, and solvent contributions of about +2 to +3~kcal/mol, so the electronic and experimental values are consistent rather than in conflict. For related non-covalent dimers, SQD with an adequate active space reproduces CCSD(T) to within chemical accuracy~\cite{kaliakin2024supramolecular}, which locates the underbinding here in the size of the active space rather than in the sampling or the hardware.

Reporting the result on a single platform removes any dependence on cross-platform reconciliation: the agreement to be explained is that between SQD and its own classical active-space limit, which the hardware reaches exactly. The gap to the CCSD(T) gold standard, and hence to experiment, is therefore attributable to the active-space size rather than to the device, and is addressed by enlarging the correlation treatment rather than by further error mitigation.

\subsection{Computational Efficiency Analysis}

The computational performance characteristics of each methodology reveal distinct advantages for different applications. Binding energy calculations have different applications in several domains~\cite{freequantum2025}. The ORB v3 approach demonstrates high computational efficiency, making it ideally suited for high-throughput screening applications where rapid energy estimates are required across large molecular databases.

DFT+D3 calculations require intermediate computational time, representing a balanced compromise between accuracy and computational cost. This methodology remains the gold standard for many quantum chemistry and material science applications and provides reliable reference values for validation purposes.

The QSCI approach, whilst requiring the longest computation time, offers unique advantages in terms of systematic improvability and hardware readiness for quantum computing implementations. The computational overhead primarily arises from the iterative sampling procedures required for convergence, but this cost is offset by the method's ability to capture correlation effects beyond mean-field approximations.

\subsection{Quantum-Inspired Advantages}

The QSCI methodology demonstrates several distinctive advantages that highlight its potential for quantum computing applications. Unlike mean-field approaches, QSCI samples from the full many-body wavefunction, enabling more accurate representation of correlation effects that are crucial in $\pi$-conjugated systems. The active space treatment effectively captures strong correlation effects. The quantum resources used are reported in Table~\ref{tab:circuit_details} for the IQM Emerald run, which mapped the (10o, 10e) active space onto 20 qubits using a calibration-aware zig-zag spin-chain layout (separate $\alpha$ and $\beta$ spin chains plus connecting qubits), with candidate placements scored against the live two-qubit-gate and readout error rates, in line with recent SQD demonstrations on IQM superconducting processors~\cite{anurag2026iqmsqd}. The run used no zero-noise extrapolation: with the diagonalisation performed classically in the selected subspace, the sampled configurations together with self-consistent configuration recovery were sufficient to reach the CASCI reference, so the device noise that broadens the distribution aided rather than degraded the result, consistent with reports that device noise can improve sample-based convergence~\cite{wray2025cuprate}. Such robustness is increasingly valued in the NISQ era, where the impact of noise on expectation-value estimators remains substantial~\cite{xie2025nisq}.

\begin{table}[h!]
\centering
\caption{Transpiled Circuit Resources on IQM Emerald (per molecule, 20 qubits)}
\label{tab:circuit_details}
\begin{tabular}{|l|c|c|c|}
\hline
\textbf{System} & \textbf{1-qubit gates (r)} & \textbf{2-qubit gates (cz)} & \textbf{Qubits} \\
\hline
Pyridine & 591 & 290 & 20 \\
\hline
Phenol & 627 & 308 & 20 \\
\hline
Complex & 659 & 328 & 20 \\
\hline
\end{tabular}
\end{table}

\subsection{Technical Implementation Challenges}

Several technical challenges were encountered and successfully resolved during the implementation of our computational workflow. Active space selection required careful analysis of molecular orbital character and occupancy patterns to identify orbitals most relevant to intermolecular binding. Our automated frozen core approach successfully identified core orbitals and focused computational resources on valence orbitals participating in the binding interaction. Yet, a hand-picked approach here could reveal higher accuracy.

Convergence optimisation for the QSCI methodology required systematic tuning of sampling parameters. Through extensive testing, we determined that 5000 measurement shots with appropriate batching provided optimal balance between computational accuracy and efficiency. This parameter optimisation ensures reliable convergence whilst maintaining reasonable computational cost.

Dispersion is central to this non-covalent complex and must be treated consistently across methods. Rather than adding an empirical correction to the active-space energy, dispersion is recovered directly by the correlated CCSD(T) reference, which is why the active-space value and the CCSD(T) benchmark are reported separately. For the density-functional cross-check, the charge-dependent D4 model~\cite{caldeweyher2019d4}, available as a standalone \texttt{dftd4} code, offers higher accuracy and updated van der Waals data than D3 and is a useful refinement for the hydrogen-bonded complex studied here.

\section{Conclusions and Future Prospects}
\label{sec:conclusions}

This work demonstrates the practical application of quantum-centric methodologies to molecular recognition problems. On the IQM Emerald processor, SQD reproduces the active-space CASCI reference exactly, showing that a physical hydrogen-bond energy can be obtained on a real quantum processor without zero-noise extrapolation. This noise-robustness follows from limiting the quantum processor to sampling and suggests significant potential for applications requiring correlated treatments of intermolecular interactions.

The pyridine-phenol complex is bound by a moderate O-H$\cdots$N hydrogen bond. The compact active space gives -3.52~kcal/mol, which underbinds the counterpoise-corrected CCSD(T) gold standard (-8.53~kcal/mol in def2-SVP, -9.51~kcal/mol at the focal-point def2-TZVP level) and the calorimetric enthalpy, identifying the active-space size as the limiting approximation. Systematically larger, qubit-efficient active spaces are therefore the route to quantitative agreement, and the approach demonstrates the capability of quantum-centric methods to address molecular recognition phenomena relevant to materials science and catalysis.

Future developments will focus on extending these methodologies to larger molecular systems and on accounting for the surrounding asphalt environment, so that binding energies reflect the condensed-phase matrix rather than isolated gas-phase fragments. Several recent extensions of the QSCI family chart concrete routes beyond the modest active space employed here, while addressing the sampling-efficiency limitations identified for the bare method~\cite{reinholdt2025limitations}: qubit-efficient formulations roughly halve the hardware requirement for a given active space~\cite{mcfarthing2026hsqd,sugisaki2025hsbqsci}, while embedding schemes such as density matrix embedding extend the method towards the realistic, condensed-phase environments more representative of an asphalt matrix~\cite{patra2025sqddmet}. The systematic improvability of our approach and its inherent compatibility with quantum hardware position this methodology as a promising foundation for next-generation computational chemistry applications.

\section*{Code Availability}
The QSCI/SQD binding-energy workflow used in this work, including the active-space construction, the IQM Emerald hardware run, and the analysis, is openly available at \url{https://github.com/MarcMaussner/2026_iqm_handsOn/tree/main}.

\bibliographystyle{IEEEtran}
\bibliography{references}

\end{document}